# Validation of the MySurgeryRisk Algorithm for Predicting Complications and Death after Major Surgery: A Retrospective Multicenter Study Using OneFlorida Data Trust

Yuanfang Ren, PhD[1,2], Esra Adiyeke, PhD[1,2], Ziyuan Guan, MS[1,2], Zhenhong Hu, PhD[1,2], Mackenzie J. Meni, PhD[1,2], Benjamin Shickel, PhD[1,2], Parisa Rashidi, PhD[1,3], Tezcan Ozrazgat-Baslanti, PhD[1,2,#], Azra Bihorac, MD, MS[1,2,#]

[1] Intelligent Clinical Care Center, University of Florida, Gainesville, FL

[2] Division of Nephrology, Hypertension, and Renal Transplantation, Department of Medicine, University of Florida, Gainesville, FL

[3] Department of Biomedical Engineering, University of Florida, Gainesville, FL

# Authors contributed equally as senior authors.

**Corresponding author:** Azra Bihorac, MD, MS, Division of Nephrology, Hypertension, and Renal Transplantation, Department of Medicine, University of Florida, PO Box 100224, Gainesville, Florida, 32610-0224, USA

Phone: 1-352-294-8580

Email: abihorac@ufl.edu

**Abstract**

**Background:** Despite advances in surgical techniques and care, postoperative complications are prevalent and affect up to 15% of the patients who undergo a major surgery. Analytic models for surgical risk estimation have the potential to benefit from the complex analysis of electronic health records and necessitates the use of sophisticated digital tools. The objective of this study is to develop and validate models for predicting postoperative complications and death after major surgery on a large and multicenter dataset, following the previously developed and validated MySurgeryRisk algorithm.

**Methods:** This retrospective, longitudinal and multicenter cohort analysis included 508,097 encounters from 366,875 adult inpatients who underwent major surgeries and were admitted to healthcare institutions within the OneFlorida+ network between 01/01/2012 and 04/29/2023. We applied the validated feature selection and transformation approach in MySurgeryRisk model, and redeveloped eXtreme Gradient Boosting (XGBoost) models for predicting risk of postoperative acute kidney injury (AKI), need for intensive care unit (ICU) admission, need for mechanical ventilation (MV) therapy, and in-hospital mortality on a development set (n=358,216 encounters) and evaluated the model performance on a validation set (n = 149,881 encounters). We presented the important features contributing to the model predictions using SHapley Additive exPlanations (SHAP) values.

**Results:** The prevalence of postoperative complications in the training and test cohorts were 8% vs 10% (n = 27,656 vs 14,646) for need of ICU admission, 4% vs 5% (n = 12,577 vs 7,858) for need of MV, 7% vs 7% (n = 25,719 vs 10,308) for AKI, and 1% vs 1% (n = 3,865 vs 1,266) for in-hospital mortality. Area under the receiver operating characteristics curve (AUROC) values were obtained for need for ICU admission, 0.93 (95% Confidence Interval [CI], 0.93-0.93); need for MV, 0.94 (95% CI, 0.94-0.94); AKI, 0.92 (95% CI, 0.92-0.92); and in-hospital mortality, 0.95 (95% CI, 0.94-0.95). Area under the precision-recall curve (AUPRC) values were computed for need for ICU admission, 0.62 (95% CI, 0.62-0.63); need for MV, 0.51 (95% CI, 0.49-0.52); AKI, 0.53 (95% CI, 0.53-0.54); and in-hospital mortality, 0.26 (95% CI, 0.24-0.29). The performance of these models is comparable to that of the previously validated MySurgeryRisk model. Primary procedure code and provider specialty consistently appeared as the top influential variables.

**Conclusion:** We developed and validated an algorithm that utilize routinely collected variables from a large multicenter cohort to predict the postoperative complications and death after major surgery, suggesting the enhanced generalizability of the model. Through our analysis, we identified critical features that significantly contribute to the predictive accuracy of the algorithm, providing valuable insights into the factors influencing surgical outcomes.

**Introduction**

Surgical procedures are among the most common medical interventions worldwide, with over 310 million major surgeries performed annually, including more than 40 million in the United States alone.[1] Despite advances in surgical techniques and perioperative care, postoperative complications remain prevalent. Complications occur in up to 15% of patients after a major surgery with reported overall postoperative mortality rate in previous research is ranging from 0.79% to 5.7%.[2] Evidence from multiple studies has shown that postoperative complications have adverse effects on patients' short- and long-term quality of life and are associated with an increased risk of reoperation, extended hospital stay and higher mortality rates.[3-5] Notably, postoperative complications not only have a greater impact on survival after major surgery than preoperative risk factors or intraoperative events, but they also impose a major financial burden with estimated additional cost of $11,626 to $19,626 per patient, primarily due to the need for intensive care unit admission or rehospitalization.[2,6,7]

Factors such as patients' age or existing comorbidities, hospital characteristics like nurse-to-patient ratio and the type and quality of operative approach as well as the anesthesia administered, have been found as key determinants for the risk of postoperative complications.[5,8] An accurate risk estimation could be useful in identifying patients who might benefit from potential strategies designed to reduce these risks. However, evaluation of these risks requires collecting and analyzing vast amounts of data found in electronic health records (EHR) which often contains critical but complex information that could be difficult to interpret using traditional methods. Aligned with widespread use of EHR, digital tools that can utilize the extensive data available to measure postoperative complication risks become increasingly crucial. Towards this aim, several postoperative complications and mortality risk calculation tools were developed.[9-11] However, these approaches either depend on information that is not readily accessible in EHRs or have not yet been adapted for integration into a clinical workflow with full automation capable of leveraging the comprehensive health data accumulated. To circumvent these challenges, we have developed and validated MySurgeryRisk platform that achieved excellent performance on predicting major complications and mortality.[12-14]

Despite being a large academic medical center, MySurgeryRisk algorithm was trained and tested on a patient cohort who were admitted to a single health institution, therefore may not represent the diversity found in broader populations. In this current study, we aim to test and validate the applicability of these findings by conducting a similar analysis using patient data curated from the OneFlorida Data Trust, with the objective of enhancing the generalizability of the outcomes. OneFlorida Data Trust, a clinical data research network comprising 14 health systems that provide care to over 20 million patients within the PCORnet Common Data Model, allows us to test and validate the surgical risk prediction model.[15,16] Its extensive collection of EHR data from a diverse patient population, coupled with a common data model different from the retrospective data used to develop the MySurgeryRisk algorithm, facilitates this process. In this work, we developed and validated models based on eXtreme Gradient Boosting (XGBoost) algorithm to predict the risk of three major postoperative complications, which are need for intensive care unit (ICU) admission, need for mechanical ventilation (MV), acute kidney injury (AKI), and in-hospital mortality.

**Methods**

***Study Design and Participants***

We obtained longitudinal EHR data for 1,455,294 hospitalized patients, encompassing 81,421,419 admissions including historical admissions, admitted to healthcare institutions within the OneFlorida+ network between January 1, 2012 and April 30, 2023. We excluded outpatient admissions, patients who were less than 18 years old as of their admission date, as well as those who did not undergo major surgery (Figure 1). Our final cohort included 366,875 encounters from 508,097 patients. Specifically, when predicting AKI complication after surgery, we excluded patients with end stage renal disease. The dataset includes demographic information, vital signs, laboratory values, medications, diagnosis and procedure codes for all admissions. The study was approved by the University of Florida Institutional Review Board and Privacy Office (IRB# IRB202300641) as an exempt study with a waiver of informed consent.

Following the guideline in Transparent Reporting of a multivariable prediction model for Individual Prognosis or Diagnosis (TRIPOD)[17] under the Type 2b analysis category, we chronologically split the cohort into training (admissions between January 1, 2012 and July 10, 2020, 70% of observations, n=358,216 encounters) and validation (admissions between July 10, 2020 and April 29, 2023, 30% of observations, n=149,881 encounters). For AKI complication, we had 348,252 encounters for training cohort and 146,458 encounters for test cohort. We used the training cohort for model development and parameter selection, and the validation cohort for model performance assessment.

### *Identification of Major Surgery and Outcomes*

We identified major surgeries using Current Procedural Terminology (CPT) codes and associated relative value units (RVUs). RVUs are a measure used in the United States healthcare system to quantify the value of medical services. Using RVUs as a guide, we selected CPT codes where the associated RVUs include an intraoperative portion denoted by an 'Intraoperative Percentage' greater than zero, and are classified as major surgery denoted by a 'Global Surgery' period of 090, indicating a 1-day preoperative period and a 90-day postoperative period within the fee schedule (Figure 2). When a patient had multiple surgeries during one admission, only the surgery with maximum intraoperative working units, calculated as 'work RVU' multiplied by 'Intraoperative Percentage', was included in the analysis. Note that since exact dates and times for the start and end of surgeries were not available, we used the date associated with the CPT code to mark the start and end dates of the surgery.

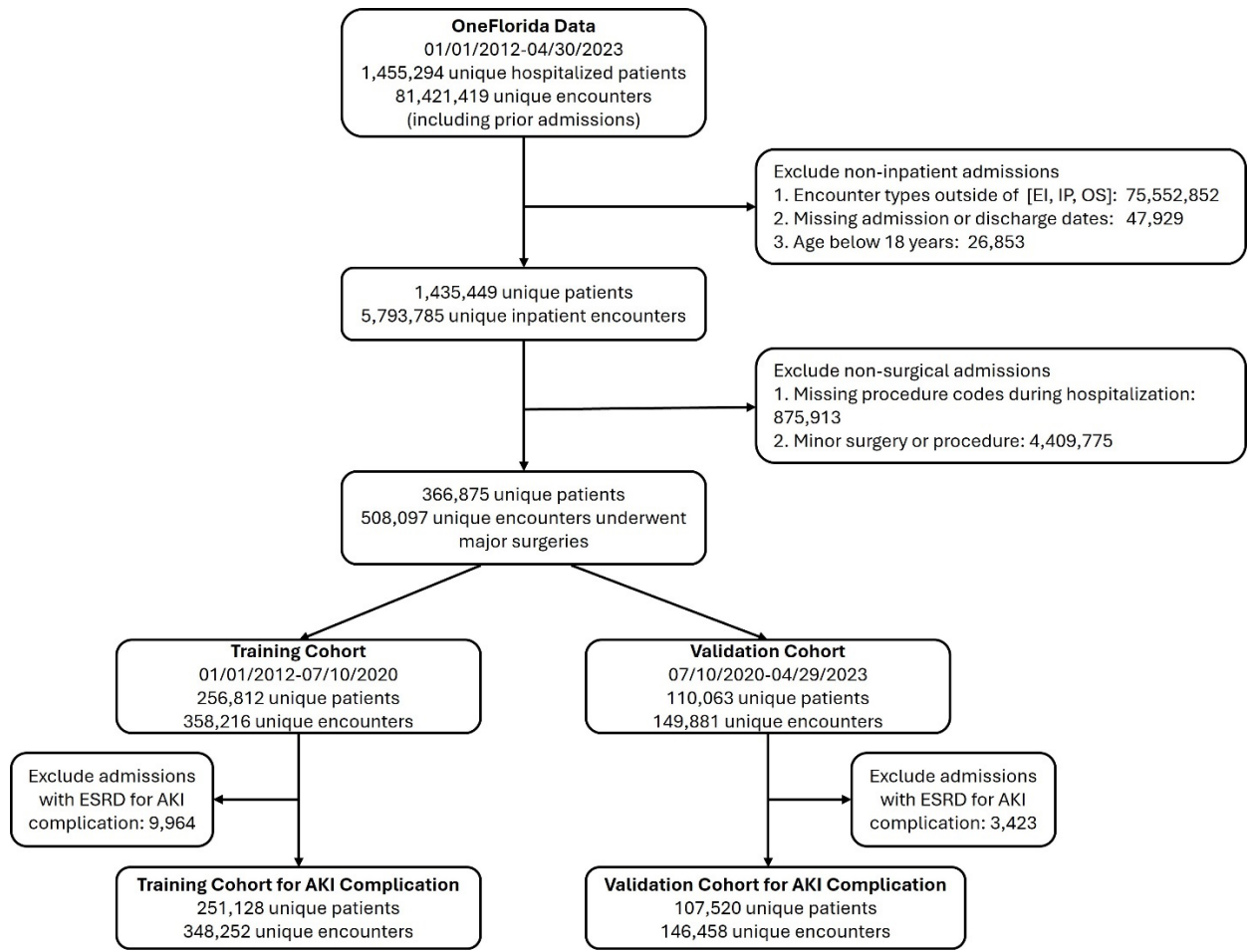

**Figure 1. Flow Diagram of MySurgeryRisk Model Development and Validation Cohorts.**
Abbreviations: EI, emergency department admit to inpatient hospital stay; IP, inpatient hospital stay; OS, observation stay; ESRD, end stage renal disease; AKI, acute kidney injury.

The primary outcomes were three postoperative complications: need for ICU admission, need for MV, AKI and in-hospital mortality. ICU admissions were identified using procedure codes, '99291', '99292', '0188T' and '0189T'. MV was identified using diagnosis codes 'V46.1', 'V46.14', 'Z99.11', 'Z99.12', 'J95.85', and 'J95.859', as well as procedure codes '96.7', '96.70', '96.71', '96.72', '5A1935Z', '5A1945Z', '5A1955Z', '94657', '94656', '1015098', '94003', '94002', '94004', and '1014859'. We utilized EHR-based computable phenotype algorithm that we previously developed and validated to automatically determine the presence of AKI[18] as per Kidney Disease: Improving Global Outcomes (KDIGO)[19] standardized serum creatinine criteria. These three outcomes were tracked from the beginning of the surgery (including the surgery date) until discharge. In-hospital mortality was determined using the date of death which were provided by institutional death records and a commercial death dataset from Datavant[15].

***Predictors***

The previously developed and prospectively validated MySurgeryRisk model at a single institution[14] utilized 137 input features, including preoperative demographic, socioeconomic, administrative, clinical, pharmacy and laboratory variables. Among the 137 features, we excluded those that were not available in the OneFlorida dataset, such as detailed surgery information (e.g., anesthesia type), those not routinely collected across institutions and thus having high missingness (e.g., smoking status and urine laboratory tests), features with high consistency (e.g., number of laboratory measurements belonging to the same laboratory panel), and features that limit generalizability of the model (e.g., doctor id and home address zip code), resulting in 99 input features. Supplemental Table 1 lists the input features and their missingness in the OneFlorida dataset.

We followed the same methods for feature processing, including data cleaning, outlier removal and missing data imputation as previously described[12,14]. We transformed categorical variables with multiple levels, such as primary procedure code, into conditional probabilities to reduce dimensionality of the data, using the log of conditional odds of positive outcome as previously described[14].

***Statistical Methods***

We applied a powerful yet simple machine learning method, XGBoost, to develop models for predicting postoperative complications. We selected the parameters 'n_estimators', 'gamma', 'max_depth', 'subsample' and 'colsample_bytree' using 5-fold cross-validation on the training cohort. After determining the optimal parameters, we retrained the model using the entire training cohort and assessed its performance on the validation cohort. We identified the important features contributing to the model predictions using SHAP (SHapley Additive exPlanations).

We assessed the robustness of models by 1) performing subgroup analysis based on sex (female vs male), race (African American vs non-African American) and age (age≤65 vs >65 years old); 2) conducting a sensitivity analysis by adding one personalized feature, doctor id, to evaluate its impact.

We evaluated each model's performance using area under the receiver operating characteristics curve (AUROC), area under the precision-recall curve (AUPRC), sensitivity, specificity, positive predictive value (PPV) and negative predictive value (NPV). We employed bootstrap sampling and non-parametric methods to obtain 95% confidence interval (CI) for all performance metrics by thresholding the probability with the value optimized Youden index. We compared clinical characteristics and outcomes of patients across cohorts using the χ2 test for categorical variables and the Mann-Whitney U test for continuous variables. The threshold for statistical significance was set at less than 0.05 for 2-sided tests. Data analysis was conducted using Python software of version 3.9, and R software of version 4.3.3.

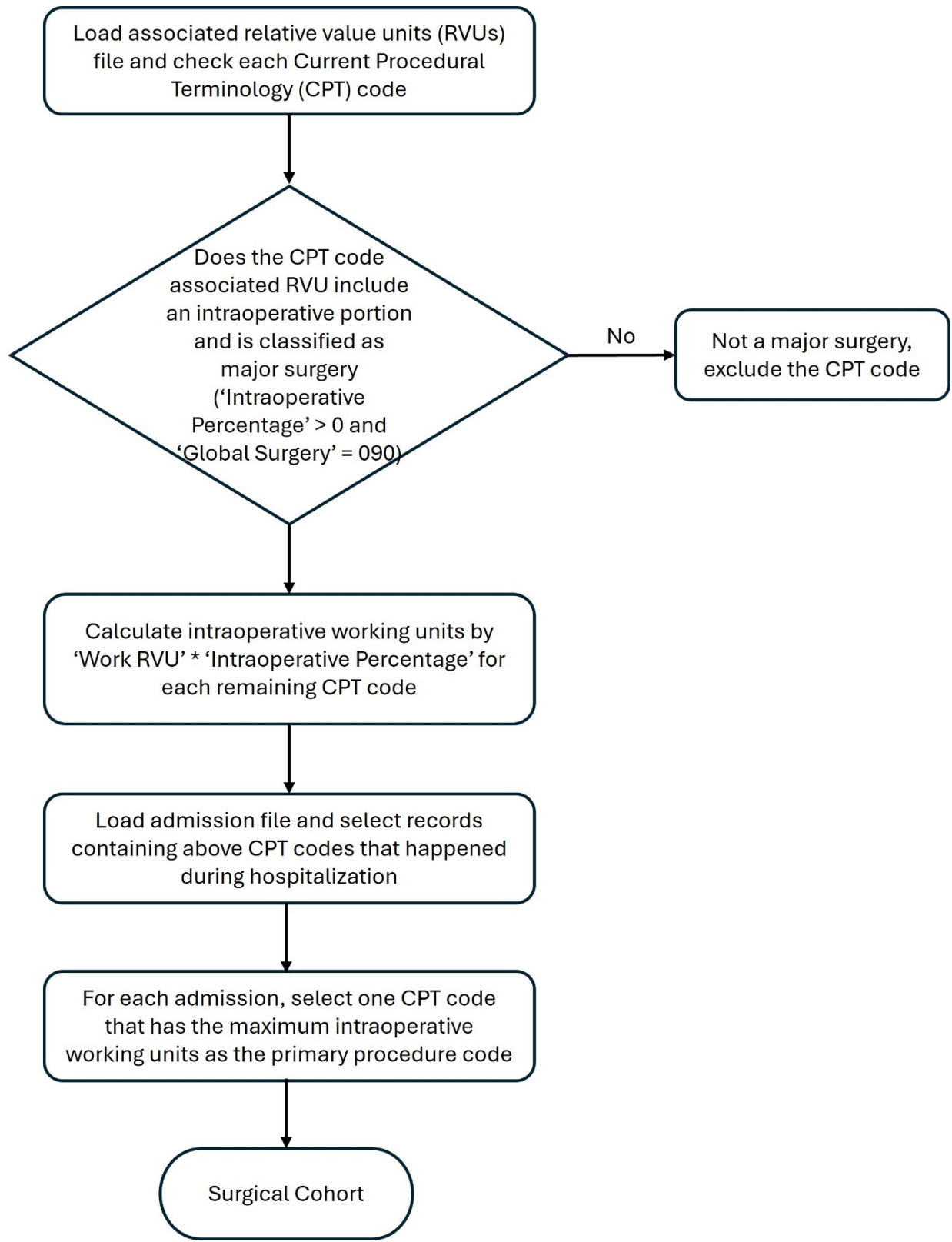

**Figure 2. Flow Diagram for the identification of major surgery and the surgical cohort**

**Results**

*Patient Baseline Characteristics and Outcomes*

Among 256,812 total patients who received 358,216 major surgical procedures in the training cohort, the mean (SD) age was 58 (19) years; 121,523 patients (53%) were female; 187,883 patients (73%) were White and 50,910 patients (20%) were African American; 45,074 patients (18%) were Hispanic; 196,145 patients (76%) had private (43%) or Medicare (33%) insurance, while the remaining 60,667 patients (24%) had Medicaid insurance (11%) or were uninsured (12%) (Table 1). The validation cohort consisting of 110,063 patients who received 149,881 major surgical procedures, exhibited statistically significant differences in demographic characteristics. Compared to the training cohort, the validation cohort had a mean (SD) age of 58 (17) years and a slightly lower proportion of female patients (50%). A slightly higher proportion of patients were White (75%), and a slightly lower proportion of patients were African American (18%). Notably, a significantly higher proportion of patients were Hispanic (26%). Additionally, there was a higher percentage of patients with private insurance (50%) or who were uninsured (14%), while the percentage of patients with Medicare (27%) or Medicaid (8%) insurance decreased.

The prevalence of postoperative complications in the training cohort was 8% for need of ICU admission, 4% for need of MV, 7% for AKI, and 1% for in-hospital mortality. There was slight variation in complication prevalence between the validation and training cohorts (e.g., in the validation cohort, the prevalence was 10% for need of ICU admission and 5% for need of MV).

**Table 1. Patient Characteristics in Training Cohorts**

| Variables | Training cohort | Validation cohort | p-value |
|---|---|---|---|
| Number of patients, n | 256,812 | 110,063 | |
| Number of surgical procedures, n | 358,216 | 149,881 | |
| Age in years, mean (SD)[a] | 58 (19) | 58 (17) | <.001 |
| Sex, n (%)[a] | | | |
|   Male | 121,523 (47) | 54,553 (50) | <.001 |
|   Female | 135,289 (53) | 55,510 (50) | <.001 |
| Race, n (%)[a,b] | | | |
|   White | 187,883 (73) | 82,915 (75) | <.001 |
|   African American | 50,910 (20) | 19,482 (18) | <.001 |
|   Other[c] | 10,105 (4) | 4,526 (4) | .01 |
|   Missing | 7,914 (3) | 3,140 (3) | <.001 |
| Ethnicity, n (%)[a,b] | | | |
|   Non-Hispanic | 203,676 (79) | 76,317 (69) | <.001 |
|   Hispanic | 45,074 (18) | 28,712 (26) | <.001 |
|   Missing | 8,062 (3) | 5,034 (5) | <.001 |
| Insurance, n (%)[a] | | | |
|   Private | 111,464 (43) | 55,047 (50) | <.001 |
|   Medicare | 84,681 (33) | 30,150 (27) | <.001 |
|   Medicaid | 29,208 (11) | 9,334 (8) | <.001 |
|   Uninsured | 31,459 (12) | 15,532 (14) | <.001 |
| Complications, n (%)[d] | | | |
|   Need for intensive care unit admission | 27,656 (8) | 14,646 (10) | <.001 |
|   Need for mechanical ventilation | 12,577 (4) | 7,858 (5) | <.001 |
|   Acute kidney injury[e] | 25,719 (7) | 10,308 (7) | <.001 |
|   In-hospital mortality | 3,865 (1) | 1,266 (1) | <.001 |

[a] Data were reported based on values calculated at the latest hospital admission.
[b] Race and ethnicity were self-reported.
[c] Other races include American Indian or Alaska Native, Asian, Native Hawaiian or Pacific Islander, and multiracial.
[d] Data were reported based on postoperative complication status for each surgical procedure. When a patient had multiple surgeries during one admission, only the surgery with maximum intraoperative working units was used in the analysis.
[e] The percentage was calculated after excluding patients with end stage renal disease. The number of surgical procedures used for postoperative acute kidney injury prediction in training and validation cohorts was 348,252 and 146,458, respectively.

### *Model Performance*

We reported models' performance in predicting postoperative complications on the hold-out validation cohort and presented the results for AUROC, AUOPRC, sensitivity, specificity, PPV and NPV with 95% CI in Table 2. AUROC values ranged from 0.92 to 0.95: need for ICU admission, 0.93 (95% CI, 0.93-0.93); need for MV, 0.94 (95% CI, 0.94-0.94); AKI, 0.92 (95% CI, 0.92-0.92); and in-hospital mortality, 0.95 (95% CI, 0.94-0.95). AUPRC values ranged from 0.26

to 0.62: need for ICU admission, 0.62 (95% CI, 0.62-0.63); need for MV, 0.51 (95% CI, 0.49-0.52); AKI, 0.53 (95% CI, 0.53-0.54); and in-hospital mortality, 0.26 (95% CI, 0.24-0.29).

**Table 2. Model performance measurements for postoperative complications with 95% confidence interval in the validation cohort**

| Complications | AUROC | AUPRC | Sensitivity | Specificity | PPV | NPV |
|---|---|---|---|---|---|---|
| Need for ICU admission | 0.93 (0.93-0.93) | 0.62 (0.62-0.63) | 0.87 (0.85-0.89) | 0.83 (0.82-0.85) | 0.36 (0.34-0.38) | 0.98 (0.98-0.99) |
| Need for MV | 0.94 (0.94-0.94) | 0.51 (0.49-0.52) | 0.87 (0.86-0.89) | 0.88 (0.86-0.88) | 0.28 (0.26-0.29) | 0.99 (0.99-0.99) |
| Acute kidney injury | 0.92 (0.92-0.92) | 0.53 (0.53-0.54) | 0.86 (0.85-0.88) | 0.8 (0.78-0.82) | 0.25 (0.24-0.26) | 0.99 (0.99-0.99) |
| In-hospital mortality | 0.95 (0.94-0.95) | 0.26 (0.24-0.29) | 0.92 (0.87-0.93) | 0.84 (0.83-0.88) | 0.05 (0.04-0.06) | 1.0 (1.0-1.0) |

Abbreviations: AUROC, area under the receiver operating characteristic curve; AUPRC, area under the precision-recall curve; PPV, positive predictive value; NPV, negative predictive value; ICU, intensive care unit; MV, mechanical ventilation.

Figure 3 presents the important features contributing to the model predictions for postoperative complications. Across all complications, primary procedure code was consistently the most important feature, indicating that the type of surgical procedure is the strongest predictor for postoperative complications. Provider specialty also consistently appeared as the second most important feature, suggesting that the expertise and specialization of the healthcare provider play a critical role in the patient's postoperative course. Comorbidity related features including fluid and electrolyte disorders and Charlson's comorbidity index, were also ranked as important predictors across complications. The feature admission source, representing whether the patient was transferred from another healthcare facility, may indicate the urgent admission and was an important predictor for all complications except for AKI. For predicting AKI, features such as age, estimated glomerular filtration rate and maximum preoperative serum creatinine were important, highlighting the relevance of renal function metrics. In predicting in-hospital mortality, additional comorbidities including coagulopathy and congestive heart failure also played significant roles.

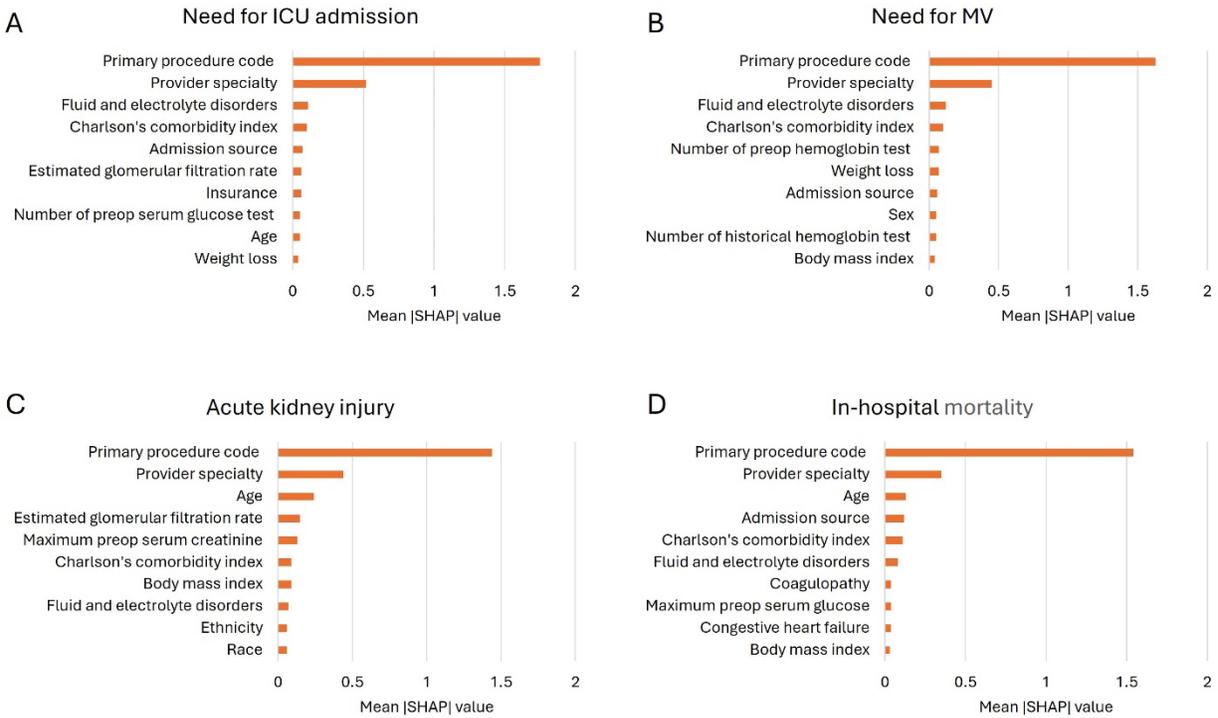

**Figure 3. Top 10 important features contributing to the prediction for postoperative complications.** Analyzed models include A) Need for ICU admissions, B) Need for MV, C) Acute kidney injury, D) In-hospital mortality. The a-axis represents the mean absolute SHAP (SHapley Additive exPlanations) values, which quantify the average contribution of each feature to the prediction model. Higher mean |SHAP| values indicate that a feature has a greater impact on the model's predictions for postoperative complications. Abbreviations: ICU, intensive care unit; MV, mechanical ventilation.

## *Robustness of Models*

Tables 3-5 present model discrimination respect to patient demographic profiles (sex, race, and age). Across all models for the four complications, the results consistently showed slightly higher AUROC, sensitivity, specificity, and NPV, but lower AUPRC and PPV for female patients compared to male patients. When comparing African American patients and Non-African American patients, the results varied across complications. For complications requiring ICU admission and MV, the model consistently demonstrated slightly better performance across all metrics for non-African American patients. In the case of postoperative AKI complications and in-hospital mortality, the model exhibited slightly higher AUROC, and sensitivity, but lower

AUPRC and PPV for non-African American patients. Model performance across age groups also varied depending on the complication. For ICU admission and MV, the model showed consistent similar and slightly higher AUROC values and higher AUPRC for patients with age less than 65. In the case of postoperative AKI complications, the model exhibited slightly higher AUROC, specificity, and NPV, but noticeably lower AUPRC and PPV for younger patients. For in-hospital mortality, the model performed slightly better across all metrics, except for PPV, in patients aged 65 and younger.

Table 3. Model performance measurements for postoperative complications with 95% confidence interval in the validation cohort stratified by sex

| Complications | Sex | AUROC | AUPRC | Sensitivity | Specificity | PPV | NPV |
|---|---|---|---|---|---|---|---|
| Need for ICU admission | Female | 0.93 (0.93-0.94) | 0.59 (0.58-0.6) | 0.87 (0.84-0.89) | 0.84 (0.82-0.87) | 0.31 (0.29-0.34) | 0.99 (0.99-0.99) |
| | Male | 0.92 (0.92-0.93) | 0.64 (0.63-0.65) | 0.86 (0.85-0.89) | 0.83 (0.8-0.84) | 0.41 (0.38-0.43) | 0.98 (0.98-0.98) |
| Need for MV | Female | 0.95 (0.95-0.95) | 0.49 (0.47-0.51) | 0.89 (0.86-0.91) | 0.88 (0.87-0.91) | 0.23 (0.21-0.28) | 1.0 (0.99-1.0) |
| | Male | 0.93 (0.93-0.94) | 0.52 (0.5-0.53) | 0.86 (0.85-0.89) | 0.86 (0.83-0.87) | 0.31 (0.28-0.32) | 0.99 (0.99-0.99) |
| Acute kidney injury | Female | 0.93 (0.93-0.93) | 0.53 (0.52-0.54) | 0.88 (0.85-0.91) | 0.81 (0.78-0.84) | 0.23 (0.21-0.26) | 0.99 (0.99-0.99) |
| | Male | 0.91 (0.9-0.91) | 0.54 (0.52-0.55) | 0.85 (0.83-0.87) | 0.79 (0.77-0.81) | 0.27 (0.25-0.29) | 0.98 (0.98-0.99) |
| In-hospital mortality | Female | 0.96 (0.95-0.96) | 0.25 (0.21-0.29) | 0.93 (0.89-0.96) | 0.87 (0.85-0.9) | 0.04 (0.04-0.05) | 1.0 (1.0-1.0) |
| | Male | 0.94 (0.93-0.95) | 0.27 (0.23-0.3) | 0.88 (0.83-0.92) | 0.85 (0.81-0.89) | 0.06 (0.05-0.08) | 1.0 (1.0-1.0) |

Abbreviations: AUROC, area under the receiver operating characteristic curve; AUPRC, area under the precision-recall curve; PPV, positive predictive value; NPV, negative predictive value; ICU, intensive care unit; MV, mechanical ventilation.

Table 4. Model performance measurements for postoperative complications with 95% confidence interval in the validation cohort stratified by race

| Complications | Race | AUROC | AUPRC | Sensitivity | Specificity | PPV | NPV |
|---|---|---|---|---|---|---|---|
| Need for ICU admission | African American | 0.91 (0.91-0.92) | 0.61 (0.6-0.63) | 0.84 (0.83-0.87) | 0.81 (0.79-0.83) | 0.35 (0.32-0.37) | 0.98 (0.98-0.98) |
| | Non-African American | 0.93 (0.93-0.94) | 0.63 (0.62-0.63) | 0.87 (0.85-0.89) | 0.84 (0.82-0.86) | 0.37 (0.35-0.4) | 0.98 (0.98-0.99) |

| | | | | | | |
|---|---|---|---|---|---|---|
| Need for MV | African American | 0.92 (0.91-0.93) | 0.5 (0.47-0.53) | 0.83 (0.81-0.85) | 0.87 (0.85-0.88) | 0.28 (0.25-0.29) | 0.99 (0.99-0.99) |
| | Non-African American | 0.95 (0.94-0.95) | 0.51 (0.5-0.52) | 0.88 (0.87-0.9) | 0.88 (0.86-0.88) | 0.28 (0.26-0.29) | 0.99 (0.99-0.99) |
| Acute kidney injury | African American | 0.9 (0.89-0.9) | 0.55 (0.53-0.57) | 0.84 (0.81-0.87) | 0.78 (0.75-0.81) | 0.29 (0.27-0.31) | 0.98 (0.98-0.98) |
| | Non-African American | 0.92 (0.92-0.92) | 0.53 (0.52-0.54) | 0.86 (0.85-0.89) | 0.82 (0.79-0.83) | 0.25 (0.23-0.26) | 0.99 (0.99-0.99) |
| In-hospital mortality | African American | 0.94 (0.92-0.95) | 0.31 (0.25-0.37) | 0.87 (0.82-0.94) | 0.86 (0.79-0.9) | 0.06 (0.04-0.08) | 1.0 (1.0-1.0) |
| | Non-African American | 0.95 (0.95-0.96) | 0.25 (0.22-0.28) | 0.92 (0.88-0.94) | 0.85 (0.83-0.88) | 0.05 (0.04-0.06) | 1.0 (1.0-1.0) |

Abbreviations: AUROC, area under the receiver operating characteristic curve; AUPRC, area under the precision-recall curve; PPV, positive predictive value; NPV, negative predictive value; ICU, intensive care unit; MV, mechanical ventilation.

**Table 5. Model performance measurements for postoperative complications with 95% confidence interval in the validation cohort stratified by age**

| Complications | Age | AUROC | AUPRC | Sensitivity | Specificity | PPV | NPV |
|---|---|---|---|---|---|---|---|
| Need for ICU admission | ≤65 | 0.93 (0.93-0.93) | 0.63 (0.62-0.64) | 0.86 (0.85-0.88) | 0.85 (0.83-0.86) | 0.37 (0.35-0.39) | 0.98 (0.98-0.98) |
| | >65 | 0.93 (0.93-0.93) | 0.61 (0.6-0.63) | 0.88 (0.85-0.91) | 0.82 (0.79-0.86) | 0.36 (0.33-0.4) | 0.98 (0.98-0.99) |
| Need for MV | ≤65 | 0.94 (0.94-0.94) | 0.52 (0.5-0.53) | 0.87 (0.85-0.89) | 0.87 (0.86-0.89) | 0.27 (0.25-0.3) | 0.99 (0.99-0.99) |
| | >65 | 0.94 (0.94-0.95) | 0.49 (0.48-0.51) | 0.88 (0.87-0.9) | 0.87 (0.85-0.88) | 0.28 (0.26-0.29) | 0.99 (0.99-0.99) |
| Acute kidney injury | ≤65 | 0.92 (0.92-0.92) | 0.52 (0.5-0.53) | 0.86 (0.83-0.89) | 0.82 (0.78-0.84) | 0.23 (0.2-0.25) | 0.99 (0.99-0.99) |
| | >65 | 0.91 (0.91-0.91) | 0.55 (0.54-0.57) | 0.87 (0.84-0.89) | 0.78 (0.77-0.81) | 0.28 (0.27-0.3) | 0.98 (0.98-0.99) |
| In-hospital mortality | ≤65 | 0.96 (0.95-0.96) | 0.3 (0.26-0.34) | 0.92 (0.89-0.94) | 0.87 (0.85-0.9) | 0.05 (0.04-0.06) | 1.0 (1.0-1.0) |
| | >65 | 0.94 (0.93-0.94) | 0.23 (0.2-0.26) | 0.89 (0.83-0.92) | 0.84 (0.79-0.88) | 0.06 (0.05-0.08) | 1.0 (1.0-1.0) |

Abbreviations: AUROC, area under the receiver operating characteristic curve; AUPRC, area under the precision-recall curve; PPV, positive predictive value; NPV, negative predictive value; ICU, intensive care unit; MV, mechanical ventilation.

To determine the impact of a personalized feature, doctor id, we conduct a sensitivity analysis by incorporating this feature into the model development and reported its performance in Table 6. Across all complications, the AUROC and NPC values for the models with and without doctor id remained nearly consistent. However, the model including doctor id consistently achieved slightly higher AUPRC and specificity, except for the complication requiring MV. On the other hand, the sensitivity was lower, except for requiring MV and AKI complications. Figure 4 presents the important features for the model including doctor id. Compared to the model without doctor id, doctor id replaced provider specialty as the second important feature for almost all complications and even ranked as the first important feature for postoperative AKI complication. Another significant difference is that, in the model without doctor id, the primary procedure code almost dominated the predictive contribution. In contrast, in the model with doctor id, the doctor id also showed a higher relative importance. The remaining important features and their ranking remained largely unchanged.

**Table 6. Sensitivity analysis adding personalized feature doctor id: model performance measurements for postoperative complications with 95% confidence interval in the validation cohort**

| Complications | With doctor id | AUROC | AUPRC | Sensitivity | Specificity | PPV | NPV |
|---|---|---|---|---|---|---|---|
| Need for ICU admission | N | 0.93 (0.93-0.93) | 0.62 (0.62-0.63) | 0.87 (0.85-0.89) | 0.83 (0.82-0.85) | 0.36 (0.34-0.38) | 0.98 (0.98-0.99) |
| | Y | 0.93 (0.93-0.94) | 0.63 (0.62-0.64) | 0.85 (0.84-0.88) | 0.86 (0.84-0.87) | 0.4 (0.37-0.41) | 0.98 (0.98-0.98) |
| Need for MV | N | 0.94 (0.94-0.94) | 0.51 (0.49-0.52) | 0.87 (0.86-0.89) | 0.88 (0.86-0.88) | 0.28 (0.26-0.29) | 0.99 (0.99-0.99) |
| | Y | 0.94 (0.94-0.95) | 0.5 (0.49-0.52) | 0.89 (0.87-0.9) | 0.86 (0.85-0.88) | 0.26 (0.25-0.28) | 0.99 (0.99-0.99) |
| Acute kidney injury | N | 0.92 (0.92-0.92) | 0.53 (0.53-0.54) | 0.86 (0.85-0.88) | 0.8 (0.78-0.82) | 0.25 (0.24-0.26) | 0.99 (0.99-0.99) |
| | Y | 0.92 (0.92-0.93) | 0.54 (0.53-0.55) | 0.87 (0.85-0.9) | 0.81 (0.79-0.83) | 0.26 (0.24-0.28) | 0.99 (0.99-0.99) |
| In-hospital mortality | N | 0.95 (0.94-0.95) | 0.26 (0.24-0.29) | 0.92 (0.87-0.93) | 0.84 (0.83-0.88) | 0.05 (0.04-0.06) | 1.0 (1.0-1.0) |
| | Y | 0.95 (0.95-0.95) | 0.26 (0.23-0.28) | 0.89 (0.87-0.93) | 0.87 (0.82-0.89) | 0.05 (0.04-0.06) | 1.0 (1.0-1.0) |

Abbreviations: AUROC, area under the receiver operating characteristic curve; AUPRC, area under the precision-recall curve; PPV, positive predictive value; NPV, negative predictive value; ICU, intensive care unit; MV, mechanical ventilation.

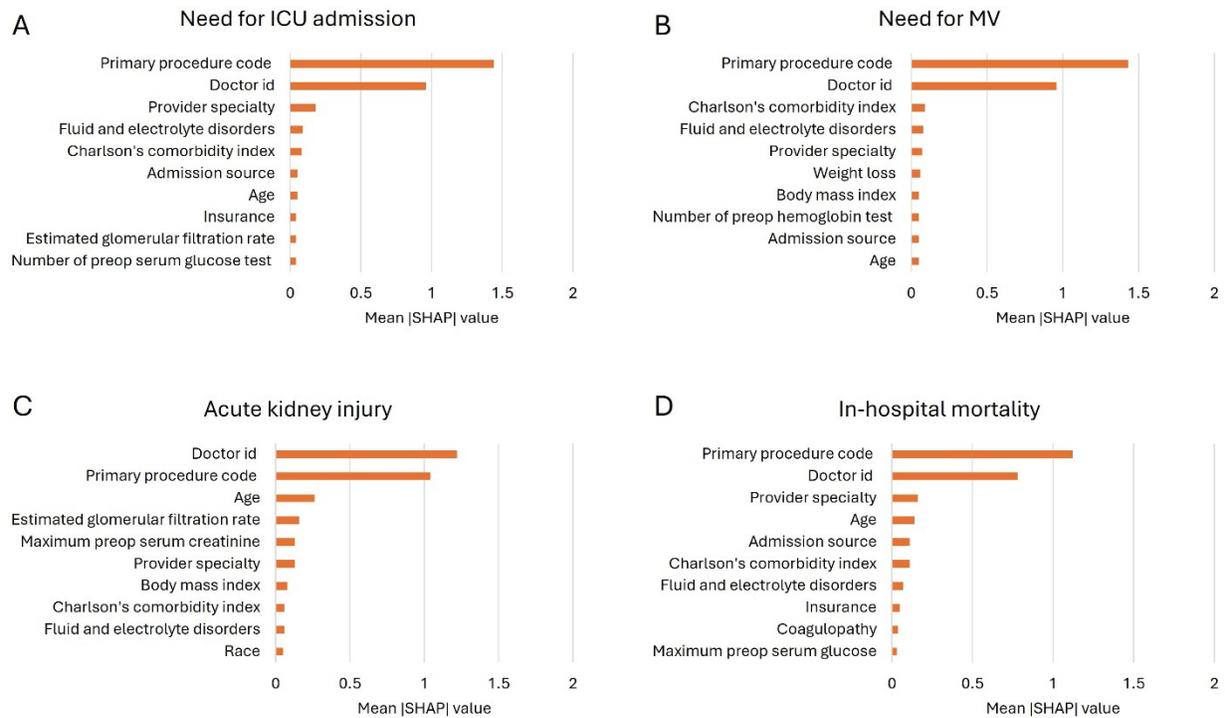

**Figure 4. Sensitivity analysis adding personalized feature doctor id: Top 10 important features contributing to the prediction for postoperative complications.** Analyzed models include A) Need for ICU admissions, B) Need for MV, C) Acute kidney injury, D) In-hospital mortality. The a-axis represents the mean absolute SHAP (SHapley Additive exPlanations) values, which quantify the average contribution of each feature to the prediction model. Higher mean |SHAP| values indicate that a feature has a greater impact on the model's predictions for postoperative complications. Abbreviations: ICU, intensive care unit; MV, mechanical ventilation.

## Discussion

In this study, we developed XGBoost models for predicting postoperative complications after major surgeries, including need for ICU admission, need for MV and AKI, and in-hospital mortality, using a multicenter dataset from OneFlorida Data Trust. We identified major surgeries using CPT codes and RUVs. Utilizing previously developed and validated MySurgeryRisk model, we carefully selected and transformed routinely collected preoperative features including demographics, comorbidities, admission information, operative information, medications and laboratory measurements for model development. The developed models achieved high AUROC values ranging from 0.92 to 0.95. Feature importance analysis demonstrated that the primary procedure code, representing the type of surgery, and provider specialty played significant roles in predicting the risk of postoperative complications.

By using only routinely collected preoperative features, the developed models have the potential to enable clinicians to assess surgical risks early, thus supporting informed decisions regarding surgery candidacy, surgical planning and preoperative care optimization. This early risk assessment capability can contribute to personalized patient management by optimizing medical conditions, tailoring surgical plans, and determining the need for additional monitoring resources such as ICU beds or specialized care teams. For example, patients identified as high risk for postoperative AKI could benefit from preoperative interventions such as fluid optimization[19], blood pressure management[19,20], medication adjustment[19], and prehabilitation programs[21,22]. By implementing these measures, healthcare providers can better prepare patients for surgery, potentially reducing the incidence and severity of postoperative complications.

The developed models following the feature selection and transformation process outlined in the previously developed and validated MySurgeryRisk[12,14], demonstrated excellent performance in terms of AUROC and validated the effectiveness of our feature selection and transformation approach. The model achieved AUROC values 0.93 (95% CI, 0.93-0.93) for the ICU admission requirement; 0.94 (95% CI, 0.94-0.94) for need for MV; 0.92 (95% CI, 0.92-0.92) for AKI; and 0.95 (95% CI, 0.94-0.95) for in-hospital mortality. This performance is comparable to the reported performance of similar complications, which are 0.88 (95% CI, 0.87-0.88) for prolonged ICU stay (greater than 48 hours); 0.91 (95% CI, 0.90-0.91) for prolonged MV (greater than 48 hours); 0.82 (95% CI, 0.81-0.83) for AKI; and 0.84 (95% CI, 0.82-0.86) for 30-day mortality.

Feature importance analysis revealed that the type of surgery, provider expertise, and patient's health condition (such as age and comorbidities) play significant roles in predicting the risk of postoperative complications, aligning findings from previous studies.[5,14,23] Dharap et al. demonstrated that the presence of comorbidities, higher surgical risk as per American Society of Anesthesiologists (ASA) grading, higher Body Mass Index, surgery type (emergency surgery, open surgery, palliative surgery, deep cavity surgery) along with other intraoperative surgical features were the significant risk factors for postoperative complications. Inclusion of personalized variables such as surgeons' previous performances in relation to his case-mix, as shown in our sensitivity analysis, demonstrated robust performance in terms of AUROC, slight performance improvement in terms of AUPRC and specificity for most complications and strong predictive power, suggesting that the local center could consider incorporating such features to reduce false positives and enhance predictive accuracy.

Subgroup analysis across patient demographic profiles revealed that the models exhibited a slight preference over female patients, non-African American patients and younger patients across most of the complications and performance metrics. Statistically significant differences in demographic characteristics and outcome prevalence between the training and validation cohorts, likely caused by data drift, may explain this issue. The validation cohort comprises samples collected during and after the COVID-19 pandemic, whereas the training cohort contains fewer samples from the pandemic period. The COVID-19 pandemic has brought significant changes to both patient lifestyles and hospital programs, potentially contributing to this observed data drift.

Our study has several limitations. First, the dataset lacks detailed information about the surgeries. As a result, identifying major surgeries and determining their exact start and end times based solely on CPT codes can introduce bias into the dataset. Additionally, the estimated surgery start and end times, coupled with the absence of other critical data such as station and respiratory device information, limit the scope of our outcomes. Consequently, the eight postoperative complications and mortality metrics in previous studies have been reduced, and prolonged ICU stay and MV (greater than 48 hours) have been redefined as mere ICU admission and MV requirements. These limitations affect the generalizability of our models. Future studies should focus on creating a multicenter surgical dataset with clear provenance, utilizing common data models, and containing abundant elements specific to surgeries as well as diverse data from multiple institutions. Second, the models lack external validation, limiting the generalizability. Third, the models exhibited bias towards certain patient groups. To prevent the creation of bias and inequalities in surgical care, future studies should aim to achieve both data and model fairness.

**Conclusions**

In this work, we validated previously developed and validated MySurgeryRisk models for predicting postoperative complications and death after major surgery on a large and multicenter dataset from OneFlorida Data Trust. We redeveloped the prediction models following the feature selection and transformation approach and achieved high and comparable performance. Further work is necessary to create a comprehensive multicenter surgical dataset surgery-specific elements, and to achieve data and model fairness.

**Supplemental Table 1. Input features used in models.**

| Feature | Model Published | Model in This Study | Type | Missingness (%) |
|---|---|---|---|---|
| **Demographics** | | | | |
| Age | X | X | Numerical | 0 |
| Sex | X | X | Binary | 0 |
| Race | X | X | Categorical | 3 |
| Ethnicity | X | X | Categorical | 3 |
| Native Language Spoken | X | X | Binary | 0 |
| Marital Status | X | | Categorical | NA |
| Smoking Status | X | | Categorical | 82 |
| Body Mass Index | X | X | Numerical | 1 |
| **Neighborhood and Socioeconomic** | | | | |
| ZIP Code of home address | X | | Categorical | 56 |
| County of home address | X | | Categorical | 56 |
| Rural or city at patient residential area | X | | Binary | 56 |
| Distance of residence to hospital, km | X | | Numerical | 56 |
| median total income at patient residential area, USD | X | | Numerical | 56 |
| prevalence of African American residents living below poverty at patient residential area, % | X | | Numerical | 56 |
| prevalence of Hispanic residents living below poverty at patient residential area, % | X | | Numerical | 56 |
| Prevalence of residents living below poverty at patient residential area, % | X | | Numerical | 56 |
| Insurance paying the bills | X | X | Categorical | 0 |
| **Comorbidities** | | | | |
| Charlson comorbidity index | X | X | Categorical | 0 |
| Myocardial Infarction | X | X | Binary | 0 |
| Congestive Heart Failure | X | X | Binary | 0 |
| Cerebrovascular Disease | X | X | Binary | 0 |
| Chronic Pulmonary Disease | X | X | Binary | 0 |
| Peripheral Vascular Disease | X | X | Binary | 0 |
| Cancer | X | X | Binary | 0 |
| Liver Disease | X | X | Binary | 0 |
| Valvular Disease | X | X | Binary | 0 |
| Coagulopathy | X | X | Binary | 0 |
| Weight Loss | X | X | Binary | 0 |
| alcohol abuse or drug | X | X | Binary | 0 |
| Fluid and electrolyte disorders | X | X | Binary | 0 |
| Chronic anemia | X | X | Binary | 0 |
| Hypertension | X | X | Binary | 0 |
| Obesity | X | X | Binary | 0 |

| Feature | Model Published | Model in This Study | Type | Missingness (%) |
|---|---|---|---|---|
| Diabetes | X | X | Binary | 0 |
| Metastatic Carcinoma | X | X | Binary | 0 |
| Depression | X | X | Binary | 0 |
| CKD status at admission | X | X | Categorical | 0 |
| Reference estimated glomerular filtration rate | X | X | Numerical | 37 |
| **Admission Characteristics** | | | | |
| Admission Source | X | X | Binary | 0 |
| Admission Day | X | X | Categorical | 0 |
| Admission Month | X | X | Categorical | 0 |
| emergent or elective admission | X | | Binary | NA |
| Yes, if the admission we happened at night | X | X | Binary | 0 |
| Medicine or surgery admitting | X | | Categorical | NA |
| **Operative Variables** | | | | |
| Current Procedural Terminology code of the primary procedure | X | X | Categorical | 0 |
| Scheduled post operation location | X | | Binary | NA |
| Scheduled room is trauma room or not | X | | Binary | NA |
| Scheduled anesthesia Type | X | | Binary | NA |
| Scheduled surgery room | X | | Categorical | NA |
| Surgery Type | X | | Categorical | NA |
| ID of Attending Surgeon | X | | Categorical | 0 |
| Specialty | | X | Categorical | 15 |
| Time from Admission to Surgery, days | X | | Numerical | NA |
| **Medication History** | | | | |
| Indicator of receiving Betablockers within one year before admission date | X | X | Binary | 0 |
| Indicator of receiving Diuretics within one year before admission date | X | X | Binary | 0 |
| Indicator of receiving statin within one year before admission date | X | X | Binary | 0 |
| Indicator of receiving Aspirin within one year before admission date | X | X | Binary | 0 |
| Indicator of receiving ACE Inhibitors within one year before admission date | X | X | Binary | 0 |
| Indicator of receiving vasopressors or inotropes within one year before admission date | X | X | Binary | 0 |
| Indicator of receiving Bicarbonate within one year before admission date | X | X | Binary | 0 |
| Indicator of receiving Antiemetic within one year before admission date | X | X | Binary | 0 |
| Indicator of receiving Aminoglycosides within one year before admission date | X | X | Binary | 0 |

| Feature | Model Published | Model in This Study | Type | Missingness (%) |
|---|---|---|---|---|
| Indicator of receiving Vancomycin within one year before admission date | X | X | Binary | 0 |
| No of nephrotoxic drugs received within one year before admission date | X | X | Binary | 0 |
| **Historical Laboratory Results** | | | | |
| Automated urinalysis, urine protein presence within 365 days prior to surgery, mg/dL | X | | Categorical | 100 |
| Automated urinalysis, urine hemoglobin within 8-365 days prior to surgery, mg/dL | X | | Categorical | 100 |
| Min hemoglobin within 8-365 days prior to surgery, g/dl | X | X | Numerical | 58 |
| Max hemoglobin within 8-365 days prior to surgery, g/dl | X | X | Numerical | 58 |
| Average of hemoglobin within 8-365 days prior to surgery, g/dl | X | X | Numerical | 58 |
| Number of hemoglobin tests within 8-365 days prior to surgery | X | X | Numerical | 0 |
| Number of urine hemoglobin tests within 8-365 days prior to surgery | X | | Categorical | 0 |
| **Preoperative Laboratory Results** | | | | |
| Automated urinalysis, urine glucose within 7 days prior to surgery, mg/dL | X | | Categorical | 100 |
| Automated urinalysis, urine hemoglobin within 7 days prior to surgery, mg/dL | X | | Categorical | 100 |
| Min hemoglobin within 7 days prior to surgery, g/dl | X | X | Numerical | 59 |
| Max hemoglobin within 7 days prior to surgery, g/dl | X | X | Numerical | 59 |
| Average of hemoglobin within 7 days prior to surgery, g/dl | X | X | Numerical | 59 |
| Variance of hemoglobin within 7 days prior to surgery, g/dl | X | X | Numerical | 59 |
| Number of hemoglobin tests within 7 days prior to surgery | X | X | Numerical | 0 |
| Min of Serum Calcium, mmol/L | X | X | Numerical | 61 |
| Max of Serum Calcium, mmol/L | X | X | Numerical | 61 |
| Average of Serum Calcium, mmol/L | X | X | Numerical | 61 |
| Variance of Serum Calcium, mmol/L | X | X | Numerical | 61 |
| Count of Serum Calcium in blood test | X | | Numerical | 0 |
| Average of anion gap in blood, mmol/L | X | X | Numerical | 64 |
| Count of anion gap in blood test | X | | Numerical | 0 |
| Min of White Blood Cell in blood, thou/uL | X | X | Numerical | 59 |
| Max of White Blood Cell in blood, thou/uL | X | X | Numerical | 59 |

| Feature | Model Published | Model in This Study | Type | Missingness (%) |
|---|---|---|---|---|
| Average of White Blood Cell in blood, thou/uL | X | X | Numerical | 59 |
| Variance of White Blood Cell in blood, thou/uL | X | X | Numerical | 59 |
| Count of White Blood Cell in blood test | X | | Numerical | 0 |
| Min of Hematocrit in blood, % | X | X | Numerical | 59 |
| Average of Hematocrit in blood, % | X | X | Numerical | 59 |
| Variance of Hematocrit in blood, % | X | X | Numerical | 59 |
| Count of Hematocrit in blood test | X | | Numerical | 0 |
| Max of Serum Red Blood Cell, Million/uL | X | X | Numerical | 59 |
| Average of Serum Red Blood Cell, Million/uL | X | X | Numerical | 59 |
| Max of the amount of hemoglobin relative to the size of the cell in blood, g/dL | X | X | Numerical | 59 |
| Average of the amount of hemoglobin relative to the size of the cell in blood, g/dL | X | X | Numerical | 59 |
| Min of Glucose in blood, mg/dL | X | X | Numerical | 60 |
| Max of Glucose in blood, mg/dL | X | X | Numerical | 60 |
| Average of Glucose in blood, mg/dL | X | X | Numerical | 60 |
| Count of Glucose in blood test | X | X | Numerical | 0 |
| Min of Serum $CO_2$, mmol/L | X | X | Numerical | 72 |
| Max of Serum $CO_2$, mmol/L | X | X | Numerical | 72 |
| Average of Serum $CO_2$, mmol/L | X | X | Numerical | 72 |
| Variance of Serum $CO_2$, mmol/L | X | X | Numerical | 72 |
| Count of Serum $CO_2$ test | X | | Numerical | 0 |
| Min of Urea nitrogen in blood, mg/dL | X | X | Numerical | 60 |
| Max of Urea nitrogen in blood, mg/dL | X | X | Numerical | 60 |
| Average of Urea nitrogen in blood, mg/dL | X | X | Numerical | 60 |
| Variance of Urea nitrogen in blood, mg/dL | X | X | Numerical | 60 |
| Count of Urea nitrogen in blood test | X | | Numerical | 0 |
| Min of Urea Nitrogen-Creatinine ratio | X | | Numerical | 86 |
| Max of Urea Nitrogen-Creatinine ratio | X | | Numerical | 86 |
| Average of Urea Nitrogen-Creatinine ratio | X | | Numerical | 86 |
| Variance of Urea Nitrogen-Creatinine ratio | X | | Numerical | 86 |
| Count of Urea Nitrogen-Creatinine ratio | X | | Numerical | 0 |
| Max of Serum Sodium, mmol/L | X | X | Numerical | 60 |
| Average of Serum Sodium, mmol/L | X | X | Numerical | 60 |
| Count of Serum Sodium test | X | | Numerical | 0 |

| Feature | Model Published | Model in This Study | Type | Missingness (%) |
|---|---|---|---|---|
| Average of Potassium in serum, mmol/L | X | X | Numerical | 60 |
| Count of Potassium in serum test | X | | Numerical | 0 |
| Max of Red cell distribution width in Blood, % | X | X | Numerical | 59 |
| Average of Red cell distribution width in Blood, % | X | X | Numerical | 59 |
| Variance of Red cell distribution width in Blood, % | X | X | Numerical | 59 |
| Min of platelet in blood, thou/uL | X | X | Numerical | 59 |
| Max of platelet in blood, thou/uL | X | X | Numerical | 59 |
| Average of platelet in blood, thou/uL | X | X | Numerical | 59 |
| Variance of platelet in blood, thou/uL | X | X | Numerical | 59 |
| Min of Serum creatinine, mg/dL | X | X | Numerical | 60 |
| Max of Serum creatinine, mg/dL | X | X | Numerical | 60 |
| Average of Serum creatinine, mg/dL | X | X | Numerical | 60 |
| Variance of Serum creatinine, mg/dL | X | X | Numerical | 60 |
| Count of Serum creatinine test | X | | Numerical | 60 |
| Max of chloride in Serum, mmol/L | X | X | Numerical | 60 |
| Average of chloride in Serum, mmol/L | X | X | Numerical | 60 |
| Variance of chloride in Serum, mmol/L | X | X | Numerical | 60 |
| Count of chloride in Serum test | X | | Numerical | 0 |